\documentstyle[tighten,floats,eqsecnum,aps, epsf]{revtex}
\draft

\begin{document}
\title{THE GEOMETRY OF THE ELONGATED PHASE IN 4-D SIMPLICIAL QUANTUM GRAVITY}
\author{Gabriele Gionti}
\address{Scuola Internazionale Superiore di Studi Avanzati\\ 
Via Beirut 2-4, 34013 Trieste, Italy \\
email gionti@sissa.it}

\maketitle

\begin{abstract}
We discuss the elongated phase of $4D$ simplicial quantum gravity
by exploiting recent analytical results. 
In particular using Walkup's theorem
we prove that the dominating configurations in the elongated 
phase are tree-like structures called "stacked spheres". Such configurations 
can be mapped into branched 
polymers and baby universes arguments are used in order to analyse 
the critical behaviour of theory in the weak coupling regime.

\end{abstract}
\vskip 2cm
\centerline{S.I.S.S.A. Ref. 144/97/FM}
\vskip 100cm
\section{Introduction}
The key concept of simplicial quantum
gravity is to work with a discrete version of space-time. To this
purpose it is necessary to consider a simplicial complex $\Sigma$ by
gluing four-dimensional simplexes through their three-dimensional faces
in such a way that the link of every
$p$- dimensional simplex $\sigma$, $p=0, 1, 2$, $link(\sigma,\Sigma)$, 
is homeomorphic to an
$(4-p-1)$-sphere. Analytically this last condition implies that  the
following Dehn-Sommerville constraints should hold 

\begin{equation}
\sum_{i=0}^{4}(-1)^{i}{N_{i}(T)}=\chi(T)   
\label{eupo} 
\end{equation}
 
\begin{equation}
\sum_{i=2k-1}^{4}(-1)^{i}{(i+1)! \over (i-2k+2)!(2k-1)!}N_{i}(T)=0  
\label{deso}  
\end{equation}

\noindent with $1\leq k \leq 2$ and $N_{i}(T)$ the number of $i$-dimensional 
simplexes of the simplicial manifold $T$ and where $\chi(T)$ is the Euler-Poincar\'e 
characteristic. Equation (\ref{eupo}) is just the Euler-Poincar\'e equation, while
(\ref{deso}) are consequence of the fact that the link of every 2k-simplex is an
odd dimensional sphere, and hence has Euler number zero.

\noindent Two simplicial manifolds $T$ and $T'$ are said to be 
PL-equivalent if there exist simplicial subdivisions of them, $\tilde{T}$ and
$\tilde{T'}$, such that $\tilde{T}$ and $\tilde{T'}$ are simplicially isomorphic.
A simplicial isomorphism \cite{Tutte} is a one to one map $\phi$ between the vertex of 
the simplicial manifolds such that $(\phi(v_i),\phi(v_j))$ is 
an edge of $T'$ if and only if $(v_i,v_j)$ is an edge of $T$ and so on 
for every simplex of any dimension. A simplicial isomorphism is an equivalence relation
among simplicial manifolds. 
An equivalence class of simplicial manifolds respect to the isomormorphism above is a PL-Manifold.

\noindent A dynamical triangulation $T_{a}$ is a simplicial manifold, 
 that is a representative  of a PL
manifold, generated by gluing, along their adjacent faces, $N_{4}(T_{a})$
equilateral simplexes, say $\sigma^{4}$, with fixed edge-length $a$. 
A dynamical triangulation gives a metric structure on a PL-manifold
(see e.g.\cite{Mauro}). 

\noindent The numbers $N_i(T)$ $i=0,...,4$ can be seen 
as the component of a five-dimensional
vector $f$

\begin{equation}
f \equiv (N_{0}(T),N_{1}(T),N_{2}(T),N_{3}(T),N_{4}(T))
\end{equation}     

\noindent called $f$-vector of the triangulation $T$. 
Dehn-Sommerville equations fix 
three component of $f$, leaving two undetermined. In simplicial gravity
$N_2(T_a)$ and $N_4(T_a)$ are, usually, the two undetermined components. 
Their values are subject to some limitations as will be explained below.

\noindent Every two dimensional simplex, in four dimention, 
is called bone (hinges in Regge calculus). 
 Let us label the bones by an index $\alpha$ and denote by $q(\alpha)$ 
the number of four simplexes
incident on it; the average number of simplexes incident on a bone is
\begin{eqnarray}
b\equiv \frac{1}{N_{2}}\sum_{\alpha}q(\alpha)=
10\left(\frac{N_4(T_{a})}{N_{2}(T_{a})} \right)
\label{avnumber}
\end{eqnarray}
since each simplex is incident on $10$ different bones. b ranges 
between two kinematical
bounds which follows from Dehn-Sommerville and a theorem
by Walkup' \cite{Walkup}. This latter theorem is also relevant in
classifying the "elongated phase" of four-dimensional simplicial quantum
gravity as we shall see below 

\noindent
{\bf Theorem}: {\it If T is a triangulation of a closed, connected
four-dimensional manifold then}

\begin{equation}
N_1(T)\geq 5N_0(T) -{15\over 2}\chi(T)
\label{importante}
\end{equation}

\noindent
{\it Moreover, equality holds if and only if $T\in H^{4}(1-{1\over 2}\chi(T))$, 
where the class of triangulations
 $H^{4}(n)$ is defined inductively according to: 
(a) The boundary complex of any abstract five-simplex ($Bd\sigma$)
is a member of $H^{4}(0)$. (b) If $K$ is in $H^{4}(0)$ and $\sigma$ is a
four-simplex of $K$, then $K'$ is in $H^{4}(0)$, where $K'$ is any complex
obtained from $K$ by deleting $\sigma$ and adding the join of the boundary
complex $Bd\sigma$ and a new vertex distinct from the vertices of $K$. (c) 
If $K$ is in $H^{d}(n)$, then $K'$ is in $H^{d}(n+1)$ if there exist two 
four-simplexes $\sigma_1$ and $\sigma_2$ with no common vertices and a 
dimension preserving simplicial
map $\phi$ from $K - {\sigma_1} - {\sigma_2}$ onto $K'$ which identifies
$Bd\sigma_1$ with $Bd\sigma_2$ but otherwise is one to one.}

\noindent 
In other words $H^{4}(0)$ is built up by gluing together
five-dimensional simplexes through their four dimensional faces and
considering only the boundary of this resulting complex. $H^{4}(n)$
differs from $H^{4}(0)$ by the fact that it has n handles. This way of
constructing a triangulation of a four-sphere has a natural connection with the
definition of a baby universe. A baby universe \cite{Jain}, \cite{1Ambjorn} 
is associated with a triangulation in which  we can
distinguish two pieces. A piece that contains the majority of the simplices
of the triangulation that is called the "mother", and a small part called the "baby".
In the "minbus" (minimum neck baby universes) the two parts are glued
together along the boundary of a four dimensional simplex (in four
dimension) that is the "neck" of the baby universe. Thus the "stacked
spheres" can be considered as a network of minbus in which the
mother is disappeared and  the babies universes have a minimal volume, that
is the boundary of a five simplices minus the simplices of the necks through
which they are glued to the others . 
We will exploit this parallel in section $VI$ to
give an estimate of the number of distinct stacked spheres. 

\noindent Establisched in \cite{Mauro}, Walkup theorem fixes a bound on the values of $b$, ,in fact  
formula (\ref{importante}) together with Dehn-Sommerville equations 
leads to $b\geq 
4-10\frac{\chi}{N_{2}}$.

\noindent From the Dehn-Sommerville equations and the fact that 
$N_{0}\geq 5$, it follows
that $b\leq 5+\frac{10\chi-50}{N_{2}}$. Thus
 in the limit of large $N$, one obtains $4\leq b\leq 5$.

\noindent As in Regge calculus the curvature 
of a dynamical triangulation is concentrated on the bones and
it depends on the number $q(B)$ of four-dimensional simplexes incident. 
 More precisely, the curvature $K(B)$ at a bone $B$  is  
\begin{eqnarray}
K(B)=\frac{48\sqrt{15}}{a^2}\left[ \frac{2\pi-q(B)\cos^{-
1}\frac{1}{4}}{q(B)} \right ]\;\;\;\; .
\label{curvapericolosa}
\end{eqnarray}

The Regge version of Einstein-Hilbert action with cosmological constant 
for a dynamical triangulation $T_a$ \cite{1Ambjorn} is 

\begin{equation}
S(T_{a})=k_{4}N_{4}(T_{a})-k_{2}N_{2}(T_{a})
\label{daction}
\end{equation}

where $k_{4}$ is
a constant depending linearly on the inverse of the 
gravitational constant and on the 
cosmological constant, whereas $k_{2}$ is proportional to the inverse of the
gravitational constant; these constants depend 
also on the edge-length $a$; typically to simplify calculations one sets $a=1$. 

The partition function for 4-D dynamical triangulations is defined as

\begin{equation}
Z(k_{2},k_{4})=\sum_{Top(M)} \sum _{N_{4}(T_{a})} \sum_{N_{2}(T_{a})}
W_{Top}(N_{2},N_{4})e^{-k_{4}N_{4}+k_{2}N_{2}}
\;\;\;\; 
\label{statmech}.
\end{equation}

\noindent where $W_{Top}(N_{4},N_{2})$ is the number of distinct {\it \'a la Tutte} 
dynamical triangulations with a
fixed number of four simplices $N_{4}$ and of bones $N_{2}$, and a fixed topology. 
Since in three and in
four dimensions 
 there is not yet a 
mathematical status for the sum over 
topologies we restrict ourselves to the simply connected topologies: 
\begin{equation}
Z_{s.c.}(k_{2},k_{4})= \sum _{N_{4}(T_{a})} 
\sum_{N_{2}(T_{a})} W_{s.c.}(N_{2},N_{4})e^{-k_{4}N_{4}+k_{2}N_{2}}
\;\;\;\; 
\label{sfera}.
\end{equation}

Equation (\ref{sfera}) has the structure of a grand canonical partition
function:
\begin{equation}
Z_{G.C.}=\sum_{N}\sum_{\sigma_{N}}e^{-\beta H(\sigma_{N})}z^{N}
\end{equation}
with the number of simplexes that is playing the role 
of the number of particles.
Following this analogy we refer to $W(N_2,N_4)$ as the 
microcanonical partition function and to
\begin{equation}
Z(k_2,N_4)=\sum_{N_2}W(N_2,N_4)e^{k_2N_2}
\label{cannolonica}
\end{equation}
as the canonical partition function. 
\vskip 1cm

\section{Canonical Partition Function}

Recently it has been analytically shown \cite{Mauro} that the dynamical
triangulations in four dimensions are characterized by two phases: a
strong coupling phase in the region $k_2^{inf}=log 9/8 < k_2 <
k_2^{crit}$ and a weak coupling phase for  $ k_2 > k_2^{crit}$.
$k_2^{crit}$ is the value of $k_2$ for which in the infinite volume
limit the theory has a phase transition from the strong to the weak
phase (for a detailed analysis see \cite{Mauro}). The 
transition between these two phases is characterized by the fact that
the subdominant asymptotics of the number of distinct
triangulations passes from an exponential to polynomial behaviour 
(c.f.\cite{Mauro}). 
Presently the precise value of $k_2^{crit}$ has not been established
yet, it is just known that it is close but distinct from $k_2^{max}=log4$, recent
numerical simulations suggest that $k^{crit}_{2}=1.24$.

\noindent In the strong coupling phase the leading term of the 
asymptotic expansion of the canonical partition function is (c.f.
\cite{Mauro}) 

\begin{eqnarray}
Z(N_4,k_2)=c_4 \left({A(k_2)+2 \over 3A(k_2)}\right)^{-4}\;\;\;
N{_4}^{-5}exp[-m(\eta^{*})N^{1\over n_H}]\nonumber\\
exp\left[[10log{{A(k_2)+2}\over 3}]N_4\right]
\label{strong}
\end{eqnarray}

where for notational convenience we have set
\begin{eqnarray}
& & A(k_{2})\equiv\nonumber\\
& & { \left [\frac{27}{2}e^{k_{2}}+1+
\sqrt{(\frac{27}{2}e^{k_{2}}+1)^2-1} \right ]}^{1/3}+
{ \left [\frac{27}{2}e^{k_{2}}+1-
\sqrt{(\frac{27}{2}e^{k_{2}}+1)^2-1} \right ]}^{1/3}-1
\end{eqnarray}
and 
\begin{eqnarray}
\eta^*(k_{2})=\frac{1}{3}(1-\frac{1}{A(k_{2})})
\;\;\;\; .
\end{eqnarray}

The explicit form of 
$m(\eta^{*}(k_2))$ and $n_H$, the Hausdorff dimension, are 
at present unknown. 

In the weak phase $ k_2 > k_2^{crit}$ the
number of distinct dynamical triangulations with equal curvature assignment have
a power law behaviour in $N_4$. The phase is caracterized by two distinct asymptotic regimes
of the leading term of the canonical partition function: the critical coupling regime and
the weak coupling regime. In the critical coupling regime 
$k_2^{crit}<k_2<k_2^{max}$ we have  

\begin{eqnarray}
Z(N_4,k_2)\asymp c_4 \left({A(k_2)+2 \over 3A(k_2)}\right)^{-4}\;\;\;
N{_4}^{\tau(\eta^{*})-5}\nonumber\\
exp\left[[10log{{A(k_2)+2}\over 3}]N_4\right]
\label{crit}\;\;\;\; ,
\end{eqnarray}
  
\noindent in which $\tau(\eta^{*})$ is not explicitly known.

Instead in the weak coupling regime $k_2>k_2^{max}$ 

\begin{eqnarray} 
Z(N, k_{2})&\asymp&\frac{c_4e}{\sqrt{2\pi}}
\frac{\eta_{max}^{-1/2}(1-2\eta_{max})^{-4}}{\sqrt{(1-3\eta_{max})(1-
2\eta_{max})}}
(\hat{N}+1)^{\tau -11/2}\nonumber \\
&&\cdot\frac{e^{(\hat{N}+1)f(\eta_{max},k_2)}}{k^{sup}_{2}-k_{2}}
\label{lower}
\end{eqnarray}

in which $\eta^{max}={1\over 4}$ for the sphere $S^{4}$ and 

\begin{equation}
f(\eta,k_{2})\equiv
-\eta\log\eta+(1-2\eta)\log(1-2\eta)-
(1-3\eta)\log(1-3\eta)+ k_{2}\eta
\label{effe}\;\;\;\; .
\end{equation}

\noindent From this form of the canonical partition function it follows that in the 
case of the sphere $S^{4}$ and in the infinite volume limit the average
value of $b$, see equation (\ref{avnumber}), is a decreasing function of
$k_2$ in the critical coupling regime and it is constantly equal to $4$ in the
weak coupling regime, that is to say 

\begin{eqnarray}
\lim_{N_4 \mapsto \infty}<b>_{N_4}={1\over\eta(k_2)},\;\;\;\; k_2^{crit}\leq 
k_2\leq k_2^{max} \nonumber\\
\lim_{N_4 \to \infty}<b>_{N_4}=4,\;\;\;\; k_2 \geq k_{2}^{max}
\label{curv}\;\;\;\; .
\end{eqnarray}

If we look at the average curvature we have

\begin{equation}
\lim_{N \to \infty} {1 \over N}<\sum_{B}K(B)vol_4(B)>
= \pi a^2 \sqrt{3}\left(10\eta^{*}(k_2) - 
{5\over \pi}cos^{-1}{1\over4}\right)
\;\;\;\;\;\;\;\;\;\; k_{2}^{crit}< k_2\leq k_{2}^{max} 
\label{vmagnetl}
\end{equation}

and

\begin{equation}
\lim_{N\to\infty}{1 \over N}<\sum_{B}K(B)vol_4(B)>
= \pi a^2 \sqrt{3} \left({5\over2} - 
{5\over \pi}cos^{-1}{1\over4}\right)\;\;\;\;\;\;\;\;\;\; 
k_2 > k_2^{max}\;\;\;\; ,
\label{vmagnetr}
\end{equation}

in which $vol_4(B)$ is the volume of the four simplexes incident on the bone $B$, and it is
$a^{2}{\sqrt{3}\over 2}q(B)$. As we have already said the value of the average curvature is 
saturated. 

\noindent This result was already known in numerical simulation
\cite{DeBakker Ph.D.} and, as we will see in detail in the next section,
it is due to the prevalence of stacked spheres in the weak coupling 
regime. 

\section{Elongated Phase}   

We have seen that in the case of triangulations of the sphere $S^{4}$
for $k_2 \mapsto log\;4$, in the infinite volume limit, $<b>_{N_4},
\mapsto 4$. Walkup's theorem \cite{Walkup} implies that the minimum
value of $b$ is reached on triangulations $K$ of the sphere $S^4$ that
belong to $H^{4}(0)$ (stacked spheres). Then for $k_2>
log\;4$  one has $<b>_{N_4}=4$,
that means that in this region of $k_2$ the statistical
ensemble of quantum gravity is strongly dominated by stacked spheres. 

\noindent As explicitly shown in \cite{Walkup},  
the elements of $H^{4}(0)$ can be put in correspondence with a
tree structure. Let us recall that a $d-$dimensional simplicial
complex $T$, $d\geq 1$, is called a simple $d-$tree if it is the closure
of its $d-$simplexes $\sigma_1,...,\sigma_t$ and these $d-$simplexes
can be ordered in such a way that: 

\begin{equation}
Cl\;\sigma_j\bigcap\left\{ \bigcup_{i=1}^{j-1} Cl\;\sigma_i \right\}= 
Cl\;\tau_j
\end{equation}

\noindent for some $(d-1)$-face $\tau_j$ of $\sigma_j$, $j\geq 2$, and
where the $\tau_j$ are all distinct. This ordering of the simplexes of $T$
induces a natural ordering of its vertices in $ v_1,...,v_{t+d}$, where
$v_{i+d}$ is the vertex of $\sigma_i$ not in $Cl\;\tau_i$. Note that the interior
part of $T$ contains the simplexes $\sigma_i$ and faces $\tau_i$. The
boundary of $T$, $Bd\;T$, consists of the boundary of the $\sigma_i$
minus the $\tau_i$, and is topologically equivalent to $S^{d-1}$. 

\noindent It can be shown\cite{Walkup}, and it is very easy to check, that any
element of $K\in H^d(0)$ is the boundary of a simple $(d+1)$-tree $T$, and
that $K$ determines uniquely the simple $(d+1)$-tree $T$ for $d\geq 2$.

Note that any stacked sphere $K\in H^{4}(0)$ can be mapped into a tree graph.
This mapping is defined in the following
way, let's consider the (unique) simple five-simple tree $T$ associated
with $K$, every five-simplex is mapped to a vertex and every
four-dimensional face in common with two five simplexes is mapped to an
edge which has endpoints at the two vertices which represent the two
five simplexes. Since the map between $K$ and $T$ is one to one, we have
a map from a stacked-sphere into a tree-graph whose number of links at
every vertex can be at most $6$ ( since a five simplex has $six$ faces).
This map, from the stacked spheres $H^{4}(0)$ to the simple tree-graphs,
is not one to one. The mathematical reason is that the previous
construction maps every $T$ to a tree-graph by an application that is
the dual map restricted to the five and four dimensional simplexes of
$T$ in its domain and whose image is the tree graphs that are the
$1$-dimensional skeleton of the dual complex. It is well known that the
dual map is a one to one correspondence only if we take in account the
simplexes of $T$ of any dimension. It follows that our map is such that
a simple tree-graph may correspond to many stacked spheres. To
illustrate better this point we shall use a picture closer to the
physical intuition. Let us consider a stacked sphere $K$ and its
simple tree $T$. Consider on $T$ one of its ordered simplexes $\sigma_j$
and let $\tau_j$ be the face that it shares with one of the simplexes
${\sigma_1,..., \sigma_t}$ introduced previously. Repeating the same
arguments used in the calculations of the inequivalent triangulations
for  baby universes \cite{Jain} , we can cut $T$ in $\tau_j$ in two
parts, such that each part has two copies of $\tau_j$ as a part of its
boundary. Since $\tau_j$ is a four dimensional simplex it is easy to
admit that we can glue this two different copies in $5\;\cdot 4\;\cdot
3$ ways to rebuild again a simple five tree $T'$. In general all the
possible gluings will generate distinct triangulations  and
consequently distinct stacked spheres (if the two parts are hightly 
simmetric triangulations some of the $60$
ways of joining will not be distinct but, since for large $N_4$ asymmetrical
triangulations will dominate, the number of case in which this will happen will be
trascurable). The corresponding tree graph associated with these configurations $T'$
will always be the same since this operation has not modified the one
skeleton of the tree $T$. 

In statistical mechanics simple-tree
structures correspond to self-avoiding "branched-polymers". So we have
that for $k_2\geq k_2^{max}$ the dominant configurations for the
triangulations of $S^4$ are branched polymers. This fact was already
observed in numerical simulations in four dimensions. 
In \cite{Ambjorn2}a network of
baby universes of minimum neck (minbu) and of minimum size (blips) 
without a mother universe was obtained. These have been
interpreted as branched polymer like structures. 

\section{Stacked Spheres and Branched Polymers}

In this section we will establish a parallel between the standard mean field 
theory of branched polymers and  the stacked spheres, in the sense
that we will use the counting tecniques of branched polymers to give an upper and
lower estimation of the number of inequivalent stacked spheres.  
As we have stressed in the previous section there exists a map between
the stacked spheres and tree graphs and this map is {\it not} one to one
in the sense that there are more stacked spheres than tree graphs. 
 This
means that the number of inequivalent stacked spheres with a fixed
number $N_5$ of five-simplexes is bounded below by the corresponding
number of tree graphs. Now we will study the statistical behaviour of
the tree graphs  using the measure of simplicial quantum gravity 
and restricting it
to the stacked spheres. In this analysis we will follow 
\cite{Frohlich} in a way adapted to our case. The reader may refer to it 
for the details. A more mathematical analysis on the
enumeration of inequivalent tree graphs is contained in \cite{Otter}. 

First of all we notice that the boundary of a five-simplex has six 
four-simplexes and every edge of a tree graph correspond to a cancellation of
two four-simplexes in the corresponding boundary of the stacked sphere.
Since in a tree graph with $N_5$ vertices there are $N_5 - 1$ edges, we 
have that the boundary of a stacked sphere, whose corresponding tree
graph has $N_5$ points, is made by a number $N_4$ of four-simplexes

\begin{equation}
N_4=4N_5 + 2
\;\;\;\; .
\end{equation}  

\noindent From the condition for stacked spheres (\ref{importante}), we
have that the Einstein-Hilbert action, for dynamical triangulations,
restricted to the stacked spheres is 

\begin{equation}
S=N_4(k_4 - {5\over 2}k_2) - 5k_2
\label{res}\;\;\;\; . 
\end{equation}

It follows that the Gibbs factor for the ensemble of tree graphs
(branched polymers) corresponding to stacked spheres is

\begin{equation}
exp\left(-(4N_5 + 2)(k_4 - {5\over 2}k_2) +5k_2 \right)
\label{gibbs}\;\;\;\; .
\end{equation}

Following the same notation of reference \cite{Frohlich}, let
$r^{6}(N_5)$ be the number of inequivalent 
rooted tree graphs with $N_5$ vertices and
with at most six incident edges on each vertex, with one rooted vertex
with one incident edge. 
$\xi^{6}(N_5)$ is the number of inequivalent tree graphs
with $N_5$ vertices. By simple geometrical considerations, or from the
asymptotic formula \cite{Frohlich} \cite{Otter}
for $r^{6}(N_5)$ and $\xi^{6}(N_5)$, we get the asymptotic formula

\begin{equation}
\xi^{6}(N_5)={1\over N_5}r^{6}(N_5 +1)
\label{asy}
\end{equation}

\noindent Now let $R^{6}(k_4,k_2)$ be the partition function for rooted
tree graphs with statistical weight (\ref{gibbs}), and $Z^{6}(k_4,k_2)$ be
the partition function for unrooted tree graphs. We have 

\begin{eqnarray}
R^{6}(k_4,k_2)&\equiv& \sum_{N_5=2}^{\infty} 
e^{-\left(4(N_5 -1)+2\right)(k_4-{5\over 2}k_2) +5k_2}r^{6}(N_5)\\
Z^{6}(k_4,k_2)&\equiv& \sum_{N_5=1}^{\infty} 
e^{-(4N_5 +2)(k_4-{5\over 2}k_2) +5k_2}\xi^{6}(N_5 )
\label{part}
\end{eqnarray}

\noindent Let's define

\begin{eqnarray}
R^{*6}(\triangle k_4)&\equiv&\left(e^{-2(k_4-5)}R^{6}(k_4,k_2)\right)
=\sum^{\infty}_{N_5=2}e^{-4(N_5-1)\triangle k_4}r^{6}(N_5)\\
Z^{*6}(\triangle k_4)&\equiv&\left(e^{-2(k_4-5)}Z^{6}(k_4,k_2)\right)
=\sum^{\infty}_{N_5=1}e^{-4N_5\triangle k_4}\xi^{6}(N_5)
\label{ric}\;\;\;\; ,
\end{eqnarray}

\noindent where $\triangle k_4\equiv k_4 - {5\over 2}k_2$.
 It is easy to see, from \ref{asy}, that 
 
 \begin{equation}
 R^{*6}(\triangle k_4)=-{1\over 4}{d\over d\triangle k_4}Z^{*6}(\triangle k_4)
 \label{pappy}\;\;\;\;.
 \end{equation}
 
 \noindent From simplicial quantum gravity it is well known that the
susceptibility $\chi(k_4,k_2)$  is given by 
 
 \begin {equation}
 \chi(k_4,k_2)\asymp {\partial^{2}\over \partial k_4^{2}} Z(k_4,k_2)
 \label{scuccu}
 \end{equation}
 
 \noindent By analogous consideration on the graph it follows that
 
 \begin{equation}
 \chi(\triangle k_4)\asymp {d^{2}\over d(\triangle k_4)^2} 
 \asymp {d\over d(\triangle k_4)} R^{*6}(\triangle k_4) 
 \label{sugg}
 \end{equation}
 
Now we will study the critical behaviour of this system in order to
obtain some information about the critical behaviour of stacked spheres. 

\noindent As easily  verified 
the following identity, which is true for rooted tree graphs 

\begin{equation}
r^{6}(N_5)=\sum_{\gamma=0}^{5}{1\over \gamma !}
\sum_{n_1,...,n_{\gamma};\sum_{i=1}^{\gamma} n_i=n-\gamma -2}
\prod_{i=1}^{\gamma}r^{6}(n_i)
\label{ram}\;\;\;\; ,
\end{equation}

\noindent  implies 

\begin{equation}
R^{*6}(\triangle k_4)=e^{-4\triangle k_4}
\left(\sum_{\gamma=0}^{5}{1\over \gamma !}
R^{*6}(\triangle k_4)\right)
\label{rec}\;\;\;\; .
\end{equation} 

\noindent Differentiating the previous 
equation we get a differential equation for $R^{*6}(\triangle k_4)$

\begin{equation}
{d\over d\triangle k_4}R^{*6}(\triangle k_4)=-{1\over 4}
R^{*6}(\triangle k_4)\left[1+{e^{-4\triangle k_4}\over
5!}\big(R^{*6}(\triangle k_4)\big)^{5}
-R^{*6}(\triangle k_4)\right]^{-1}
\label{diffe}
\end{equation}

\noindent So $R^{*6}(\triangle k_4)$ shows a singularity when
\begin{equation}
R^{*6}\left((\triangle k_4)^{c}\right)=1+{e^{-4\triangle k_4}\over
5!}\left(R^{*6}\big((\triangle k_4)\big)^{c}\right)^{5}
\label{crite}
\end{equation}

\noindent This equation has only one real solution 
and since ${d\over d\triangle k_4}R^{*6}|_{(\triangle
k_4)^{c}}$ diverges, the inverse function is zero 

\begin{equation}
\left.{d(\triangle k_4)(R^{*6})\over dR^{*6}}\right|_{(\triangle k_4)^{c}}=0
\label{inver}
\end{equation}

\noindent By the implicit function theorem $(\triangle k_4)(R^{*6})$ is
analytic function at $R^{*6}\left((\triangle k_4)^{c}\right)$, which
implies 

\begin{eqnarray}
\triangle k_4(R^{*6}) - {\triangle k_4}^{c}&=&
{1\over 2}\left.{d^{2}(\triangle k_4)(R^{*6})
\over d(R^{*6})^{2}}\right|_{R^{*6}\left((\triangle k_4)^{c}\right)} 
\left(R^{*6}(\triangle k_4) - R^{*6}({\triangle k_4}^{c})\right)^{2}\nonumber\\
& &+ o\left(\left(R^{*6}(\triangle k_4) - R^{*6}({\triangle k_4}^{c})\right)^{2}\right)
\label{exp}\;\;\;\; .
\end{eqnarray}

\noindent So near $(\triangle k_4)^{c}$ we can write

\begin{equation}
R^{*6}(\triangle k_4)\asymp R^{*6}\big((\triangle k_4)^{c}\big)
+ C\left((\triangle k_4) - (\triangle k_4)^{c}\right)^{1\over 2}
\label{svilu}
\end{equation}

\noindent From (\ref{sugg}) we get

\begin{equation}
\chi^{6}(\triangle k_4)\asymp
\left((\triangle k_4) - (\triangle k_4)^{c}\right)^{-{1\over 2}}
\label{scippu}
\end{equation}

\noindent This means that the critical exponent of the susceptibility
for the system of these tree graphs is $\gamma={1\over 2}$.

The partition function that has been studied is a lower estimate of the
partition function of the stacked spheres. Consider, now, a stacked
sphere $K$ and a face $\tau_j$ through which two five-simplexes are
glued together (a link on the corresponding tree graph). We can glue a
two four face of a stacked sphere in $5{\cdot} 4{\cdot} 3$ different
ways, for large number of simplexes this will generate distinct
configurations of stacked spheres whose associated tree-graph is always
the same. Repeating the same argument for every $j$, $j=1,...,N_5-1$,
(that is to say for every link of the dual tree graph) we obtain a
factor, $(5{\cdot} 4{\cdot} 3)^{N_5-1}$, that multiplied by
$r^{6}(N_5)$ and $\xi^{6}(N_5)$, gives an upper bound on the number of ,
respectively, rooted and unrooted inequivalent stacked spheres. In other
words 

\begin{equation}
Z_{tree}(k_4,k_2)\leq Z_{s.s.}(k_4,k_2) \leq {\tilde Z}_{tree}(k_4,k_2)
\label{dise}
\end{equation} 

\noindent where $Z_{tree}(k_4,k_2)$ is the partition function for tree
graphs studied above, $Z_{s.s.}(k_4,k_2)$ is the partition function for
stacked spheres and ${\tilde Z}_{tree}(k_4,k_2)$ is the partition
function for tree graphs with the additional weight defined above.

\noindent A similar analysis as for $Z_{tree}(k_4,k_2)$ shows that the
critical line of ${\tilde Z}_{tree}(k_4,k_2)$ is, of course, a straight
line parallel and above respect to the $k_4$ axis to $Z_{tree}(k_4,k_2)$
one, and the susceptibility exponent is again $\gamma ={1\over 2}$. 

\noindent Obviously the estimates (\ref{dise}) are true for the canonical
partition functions too, 

\begin{equation}
Z_{tree}(N_4,k_2)\leq Z_{s.s.}(N_4,k_2) \leq {\tilde Z}_{tree}(N_4,k_2)
\label{estimo}\;\;\;\; ,
\end{equation}

\noindent and the previous calculations show that

\begin{equation}
Z_{tree}(N_4,k_2)\asymp N_{4}^{-{5\over 2}}
e^{-N_4(k_4 -{5\over 2}k_2 -t_{4})}
\label{estimo1}
\end{equation}

and

\begin{equation}
{\tilde Z}_{tree}(N_4,k_2)\asymp N_{4}^{-{5\over 2}}
e^{-N_4(k_4 -{5\over 2}k_2 +t_{4}-{1\over 4}log\;60)}
\label{estimo2}\;\;\;\; ,
\end{equation}

\noindent where $t_4$ is a constant that may be calculated by the 
equation \ref{crite}. More easly from the table in reference \cite{Otter}
we get that $t_4 \approx {1\over 4}log\;0.34$. 

\noindent The circumstance that in the weak coupling region 
the partition function of
quantum gravity, as found in an analytically way in \cite{Mauro}, is
strongly dominated  by stacked spheres, allows us to write 

\begin{equation}
Z_{s.s.}\asymp N_4^{\gamma_s - 3}e^{k_4^{c}(k_2)}
\label{funct}\;\;\;\; ,
\end{equation}

\noindent where $\gamma_s$ is the susceptibility exponent \cite{Ambjorn}.

\noindent 
Thus the critical line of the stacked spheres is a straight line parallel and 
between the critical lines of the systems of the two branched polymers. This 
implies

\begin{equation}
 k_4 -{5\over 2}k_2 +t_{4} \leq k_4^{c}(k_2) \leq 
 k_4 -{5\over 2}k_2 +t_{4}-{1\over 4}log\;60
\label{estimo3}
\end{equation}

Moreover from the equations \ref{estimo1} and \ref{estimo2} 
the one loop green functions \cite{1Ambjorn} 
of the two model
of 
branched polymers have resptectivly , near their critical lines ,  
the asymptotic behaviour

\begin{equation}
G_{tree}(\triangle k_4)\asymp cost_{1} + (\triangle k_4 - {\triangle k_4}^{c})^{1\over 2}
\;\;\;\;\;\;\;
{\tilde G}_{tree}(\tilde{\triangle k_4})\asymp
cost_{2} + (\tilde{\triangle k_4} - {\tilde{\triangle k_4}}^{c})^{1\over 2}
\label{1green}
\end{equation}

This last equation togheter the equations \ref{estimo} and \ref{funct}
prove that the one loop function of the stacked spheres $G_{s.s.}(k_4,k_2)$
near the critical line has the asymtotic behaviour

\begin{equation}
G_{s.s.}(k_4,k_2)\asymp cost_{3} +(k_4 -k_4^{c}(k_2))^{1-\gamma_s}
\label{sgreen}\;\;\;\; ,
\end{equation}

\noindent with $\gamma_s < 1$ (the value $\gamma_s=1$ is not allowed because in this case 
the one loop green function of stacked spheres at the critical line will have a behaviour
like $\sum_{N_4=5}^{\infty}1/N$ that is divergent and then incompatible with the upper bound 
given by the second equation of \ref{1green}).

Motivated by physical considerations, we can use a well known argument
\cite{1Ambjorn} \cite{LJ} in favour of the fact that the susceptibility exponent
of the stacked spheres is $\gamma_s={1\over 2}$. 
More precisely we will show that a model of proliferating baby
universes, with the measure of quantum gravity restricted to stacked
spheres, can be put in correspondence with the statistical system of
stacked spheres. 

\noindent Let us consider four dimensional triangulations that are
(boundary of the) stacked spheres in which there can be loops made by
two three-dimensional simplexes. This is possible whenever the stacked spheres
are pinched on a three simplex creating a bottle neck loop of two
three-simplexes. These loops could be either  the loops of a Green
function or the minimum bottle neck of a baby universes. This class of
triangulations, following the notation in literature, is called
${\bf T}_{2}$. The other class of triangulations is the stacked spheres
in which the minimum loop length can be made by the boundary of a
four-simplex that are five three-simplexes. We call this last class ${\bf T}_5$. 
In the two dimensional theory, the introduction in ${\bf T}_{2}$ of
two-link loops (the two dimentional analogous of two three-simplex loop)
corresponds in the matrix model $\phi^{3}$ to consider Feynman diagrams
with self-energy (c.f. \cite{1Ambjorn} \cite{LJ}). 

\noindent Since the minima loops of ${\bf T}_2$ and ${\bf T}_5$, for
which they differ, are of the order of lattice spacing we will expect
that the two classes of triangulations, as a statistical
mechanics system, coincide in the scaling
limit, that is to say they belong to the same universality class. 
  
\noindent Let's consider the minimum neck one loop function $G(\triangle
k_4)$ in ${\bf T}_2$. In every triangulation of ${\bf T}_2$ we can cut
out the maximal size baby universe of minimum neck and close the two
three-simplex loop. We will obtain again a triangulation that belongs to
${\bf T}_2$. In this way we will obtain all the triangualtions of ${\bf
T}_2$ from the triangulations of the stacked spheres ${\bf T}_5$
considering that for each three-simplex either leave them in their
actual form or we can open the triangulation to create a two
three-simplex loop and gluing on it a whole one loop universe
$G(\triangle k_4)$.  We note that in the triangulations of ${\bf T}_5$
$\overline {N}_3=5/2\overline {N}_4$ (Dehn-Sommerville).  Calling the
one loop function  of ${\bf T}_5$ $\overline{G}(\overline{\triangle
k_4})$, the above considerations lead to the identity 
 
 \begin{equation}
 G(\triangle k_4)=\sum_{{\bf T}\in {\bf T}_5}
 e^{-\overline{N}_4\triangle k_4 }
 (1+G(\triangle k_4))^{\overline{N}_3}=
 \sum_{{\bf T}\in {\bf T}_5}e^{-\overline{N}_4\overline{\triangle k_4}}
 =\overline{G}(\overline{\triangle k_4})
 \label{selfenerg}\;\;\;\; ,
 \end{equation}
 
\noindent in which $\triangle k_4=k_4 - 5/2 k_2$ comes out from restricting the 
 action of quantum gravity to stacked spheres \ref{res}, and where we have 
 defined
 
 \begin{equation} 
 \overline{\triangle k_4}=\triangle k_4 - 
 {5\over 2}log\left(1+G(\triangle k_4)\right)
 \label{ini}\;\;\;\; .
 \end{equation}

\noindent By \ref{selfenerg} we can also write last equation as  

\begin{equation} 
 \triangle k_4=\overline{\triangle k_4} + 
 {5\over 2}log\left(1+\overline{G}(\overline{\triangle k_4})\right)
 \label{idi}\;\;\;\; .
 \end{equation}

\noindent By universality and the estimates (\ref{estimo}) it follows that near the 
critical point we have that $G(\triangle k_4)\asymp cost+(\triangle k_4 -
\triangle k^{c}_4)^{1-\gamma_s}$ with $\gamma_s < 1$. 

\noindent The susceptibility functions of ${\bf T_2}$ and ${\bf T}_5$ by \ref{selfenerg}
are
\begin{equation}
\chi(\triangle k_4)\asymp -{d \over d\triangle k_4}G(\triangle k_4)\;\;\;\; 
\overline{\chi}(\overline{\triangle k_4})\asymp -{d \over d\overline{\triangle k_4}}
\overline {G}(\overline{\triangle k_4})
\end{equation}

\noindent From \ref{idi} we have

\begin{equation}
{d(\triangle k_4)\over d(\overline{\triangle k_4})}
=1-{5\over 2}
{\overline{\chi}(\overline{\triangle k_4})\over (1+\overline{G}(\overline{\triangle k_4})}
\label{travo}\;\;\;\; .
\end{equation}

\noindent If we calculate di derivative with respect to $\triangle k_4$
of the one loop function $G(\triangle k_4)$ and use the previous
equation we get 

\begin{equation}
\chi(\triangle k_4)={\overline{\chi}(\overline{\triangle k_4})
\over {1-{5\over 2}
{\overline{\chi}(\overline{\triangle k_4})\over (1+\overline{G}(\overline{\triangle k_4})}}}
\label{rela}\;\;\;\; .
\end{equation}

Now it is clear that $\chi(\triangle k_4) \mapsto +\infty$ for $\triangle k_4 \mapsto 
(\triangle k_4)^{c}$ and with the same critical exponent 
$\overline{\chi}(\overline{\triangle k_4)} \mapsto +\infty$ for 
$\overline{\triangle k_4} \mapsto (\overline{\triangle k_4})^{c}$. 
From equation (\ref{rela}) when  $\chi(\triangle k_4) \mapsto +\infty$ we have that 
$\overline{\chi}\left(\overline{\triangle k_4}\right)\mapsto
 {2\over 5}\Bigl(1+\overline{G}\left(\overline{\triangle k_4}
({\triangle k_4}^{c})\right)\Bigr) < +\infty$, 
then the system ${\bf T}_5$ is above his critical line, i.e.
$\overline{\triangle k_4}({\triangle k_4}^c)>\overline{\triangle k_4}^{c}$. 
These facts  
imply that at $\overline{\triangle k_4}({\triangle k_4}^{c})$ 
$d(\triangle k_4)/d(\overline{\triangle k_4})=0$ and arount it 
$\overline{\chi}(\overline{\triangle k_4})/
(1+\overline{G}(\overline{\triangle k_4}))$ 
is a decreasing monotonic
function  
by equation \ref{travo} because $\chi(\triangle k_4) \mapsto +\infty$ 
, we can expand equation \ref{idi} and \ref{rela} 
around $\overline{\triangle k_4}({\triangle k_4}^{c})$ 

\begin{equation}
\triangle k_4 -{\triangle k_4}^{c}= 
cost\left(\overline{\triangle k_4} -\overline{\triangle k_4}({\triangle
k_4}^{c})\right)^{2}
\label{solita}\;\;\;\; .
\end{equation}

\begin{equation}
\chi(\triangle k_4)\asymp
{1\over {\overline{\triangle k_4} -\overline{\triangle k_4}({\triangle k_4}^{c}})}
\asymp {1\over \sqrt{\triangle k_4 - {\triangle k_4}^{c}}}
\label{final}\;\;\;\; .
\end{equation}

\noindent The  second asymptotic equality of the last equation implies

\begin{equation}
\gamma_s={1\over 2}
\label{provola}
\end{equation}

The dominance of stacked spheres in the weak phase allows us 
to fix the parameter $\tau$ in the partition function 
of quantum 
gravity in the weak coupling regime.

\begin{equation}
\tau -{11\over 2}=-{5\over 2} \Longrightarrow \tau=3
\label{crippu}
\end{equation}     

\acknowledgments

It is a pleasure to thank J. Ambj{\o}rn, M. Carfora, M. Caselle, A. D'Adda and D. Gabrielli
for interesting discussions. In particular G. Esposito and C. Stornaiolo for
careful reading the manuscript.

\end{document}